\documentclass[rnote,traditabstract]{aa} 
\usepackage{graphicx}
\usepackage{txfonts}
\usepackage{natbib}
\def \lya\ {Lyman-$\alpha\,$}
\begin{document}
   \title{Uncovering the magnetic environment of our solar system}
\titlerunning{Magnetic environment of our solar system}

   \author{L. Ben-Jaffel
          \inst{1, 2}
          \and
          R. Ratkiewicz \inst{3,4} }

   \institute{UPMC Univ Paris 06, UMR7095, Institut d'Astrophysique de Paris,
              F-75014 Paris, France. \email{bjaffel@iap.fr}
         \and
               CNRS, UMR7095, Institut d'Astrophysique de Paris, F-75014, Paris, France
         \and
              Institute of Aviation, Warsaw, Poland
         \and
             Space Research Center, Polish Academy of Sciences, Warsaw, Poland. 
             \email{roma@cbk.waw.pl} }

   \date{Received May 21, 2012; accepted August 11, 2012}

 
\abstract{Since its formation 4.6 billion years ago, our solar system
has most likely crossed numerous magnetized interstellar clouds and bubbles of
different sizes and contents on its path through the Milky Way. Having a
reference model for how the heliosphere and interstellar winds interact is
critical for understanding our current Galactic environment, and
it requires untangling the roles of two major actors: the time-variable
solar wind and the local interstellar magnetic field. Numerical
simulations predict a distortion of the heliosphere caused by both solar
wind anisotropy and interstellar magnetic field orientation. However,
model comparison to deep space probes' measurements led to contradictory
reports by {\it Voyager} 1 and {\it Voyager} 2 of both several crossings
of the solar wind's termination shock and of the strength of the local
interstellar field, with values ranging from 1.8 to 5.7 $\mu$G. Here, we
show that {\it Voyager} 1 \& 2 plasma, fields, and \lya\ sky background
measurements, as well as space observations of high-energy particles of
heliospheric origin, may all be explained by a rather weak interstellar
field $2.2 \pm0.1\, \mu$G pointing from Galactic coordinates 
$(l,b)\sim(28, 52)\pm3^\circ$. For the 2000 epoch Ulysses-based helium
parameters assumed thus far, the interstellar bow shock must exist. 
By contrast, using the 2010 epoch IBEX-based He parameters and 
a stronger magnetic field
 leads to a plasma configuration that is not consistent with the
 Voyagers TS crossings.  For the newly proposed interstellar He parameters,
more simulations are required before one may determine whether the
interstellar bow shock truly does disappear under those assumptions.}  

\maketitle
%

\section{Introduction} 
The supersonic, expanding solar wind (SW) slows through the termination
shock (TS) to adjust to the outer conditions and then carves a cavity into the local
interstellar medium (LISM), which is called the heliosphere . While the asymmetry 
of the heliosphere caused by the local interstellar
magnetic field (ISMF) was first predicted by the Newtonian approximation
\citep{fahr88}, it was not confirmed until 10 years later by full 3D magnetohydrodynamic (MHD) 
theoretical models (e.g., \cite{rat98}). Since then, modelers quickly
realized that the observation/modelling of such asymmetry may
constrain the direction of the ISMF and act as an interstellar magnetic
compass. Before MHD models were employed, the first attempt to uncover
the ISMF vector was made using observations of an asymmetry in the
spatial distribution of the \lya\ emission far from the Sun \citep{bpr00}. 
Assuming the observed \lya\ glow of heliospheric origin, the
simple Newtonian approximation showed that the ISMF vector points 
$\sim 40^{\circ}$ away from the LISM flow direction with a strength of $\sim 1.8\, \mu $G. 
These results were later confirmed by MHD 3D simulations, providing 
an independent estimation of the ISMF direction pointing from Galactic
coordinates $(37\pm 14^{\circ}, 49.5 \pm 8.5^{\circ} $) \citep{rbg08}. 

Apart from the \lya\ radiation constraints on the heliosphere models, it is
commonly believed that {\it Voyager} 1 (V1) crossed the TS for the first time
at the heliocentric distance $\sim94$ AU on 2004 December 16 
\citep{bur05,dec05,gur05,sto05}, and {\it Voyager} 2 (V2) at $\sim84$ AU on
2007 August 30 \citep{bur08,dec08,gur08,rich08}. The 10 AU difference 
in the crossing distance indicates heliospheric asymmetries that 
could result from the ISMF, the asymmetric SW dynamic pressure, or the motion of TS caused 
by any time-dependent phenomena \citep{sto08}.
\begin{table*}
\caption{Summary of models indicating the strength and orientation of the ISMF.}
\begin{center}
\scalebox{.95}{
\begin{tabular}{ccccccccc}
\hline
ISMF strength ($\mu$G)&  ISMF orientation    & B$_SW$&  Solar-wind  &    \lya\    &   V1   &  V2   &	IBEX   &   Ref. \\
    1.8           &  $\alpha\sim40^\circ$    &   -   &     SS       &       +     &   -    &  -    &      -    &    \citep{bpr00}   \\ 
     -            &   $\beta=0$              &   -   &     -        &       +     &   -    &  -    &      -    &     \citep{lal05}   \\
    1.8	         &($228\pm14^\circ, 42\pm8.5^\circ$)&	 -   &	   SS       &       +     &   -    &  -    &      -    &    \citep{rbg08}   \\
    3.8	          & (248, 35)$\pm5^\circ$    &   -   &     SS       &       -     &   +    &  +    &      -    &    \citep{rg08}   \\
  3.7-5.5  & $\alpha\sim20-30$, $\delta<90$  &   ?   &     SS       &       -     &   +    &  +    &      -    &     \citep{oph09}   \\	
    $>4.0$	  & $\alpha\sim30$, $\beta=0$ &	 +   &     SS       &       -     &   +    &  +    &      -    &    \citep{pog09}   \\	
    3.0	          &      (224, 41)$^\circ$   &   ?   &     SS       &       -     &   -    &  -    &      +    &     \citep{heer10}   \\	
   4.4	          &    $\alpha\sim20$        & 	 ?   &     SS       &       -     &   -    &  -    &      +    &    \citep{chal10}   \\
   3.0$\pm$1.0        & (225, 35)$\pm5^\circ$    &   -   &     SS       &       -     &   -    &  -    &      +    &     \citep{gryg11}   \\
  2.4$\pm$0.3     & (227, 35)$\pm7^\circ$    &   +   &     NS       &       -     &   -    &  +	   &      +    &    \citep{stru11}   \\	
   2-3	          & (220-224, 39-44)$^\circ$ &   +   &     SS       &		 +     &   -    &  -    &      +    &     \citep{heer11}\\	
  $\geq 3.0$           & $\alpha=45^\circ$         &   +   &     NS       &  - & - & - & + &  \citep{mcco12}\\
  2.2$\pm0.1$   &  (224, 36)$\pm3^\circ$	  &   +   &	   NS	     & 	    +     &   +    &  +    & 	  +    &   This work   \\
\hline
\end{tabular} }
\end{center}

\tablefoot{$\alpha$ is angle between inflow $\mathbf{V}_{\mathbf{\infty}}$ and magnetic 
field $\mathbf{B}_{\mathbf{\infty}}$ vectors. $\beta$ is angle from HDP
 plane. { For all references except McComas et al. (2012), the HDP was
 defined  by Lallement et al. (2010). The ISM direction, speed, plasma
 temperature, and HDP orientation have different values in  McComas et al. (2012)}. Coordinates are ecliptic. B$_{SW}$ is the interplanetary 
magnetic field. $\delta$ is the angle between the BV plane and the Sun's
 equatorial plane (noted $\beta$ in Opher et al. (2009)). 
\lya\ corresponds either to Voyagers or to SOHO/SWAN ultraviolet observations. V1 and V2 correspond to plasma and field {\it in situ} 
measurements. IBEX corresponds to the ENA ribbon observations. SS is for spherical symmetric solar wind flow and 
NS is for non-spherical symmetric.\label{tbl-1}}
\end{table*}

More recently, {\it IBEX} has completed sky maps 
that image energetic neutral atoms (ENAs). These maps reveal a bright
ribbon of ENAs in the energy range $\sim 0.2-6$ keV 
\citep{mcco09, funs09, fuse09, schw09}. We note that
 {\it Cassini} observed $\sim$6-13 keV ENAs with a similar but
broader structure \citep{krim09}. One possible explanation assumes
that the ribbon discovered by IBEX is probably ordered by the ISMF
interacting with the heliosphere \citep{mcco09, heer10,chal10}. Because they
provide global maps of the interstellar interaction, ENAs observations
are highly complementary to and synergic with the detailed single {\it in
situ} measurements provided by the {\it Voyager} probes. 

Most of these observations as well as further measurements of the plasma flow
and magnetic field in the inner heliosheath by {\it Voyager} 2, have spurred numerous
attempts to determine the parameters of ISMF from MHD and neutrals
modeling. Table 1 provides a list of papers in which the ISMF
parameters have been estimated. A quick look shows that up to now no analysis 
thus far proposed has used the full set of observations. In fact, with the last test 
reported, it was not possible to get the ISMF strength and
orientation consistently using the crossing TS distance from the Sun for
V1 $\sim94$ AU, for V2 $\sim84$ AU, and for the IBEX ribbon angular
location for the same initial SW conditions \citep{stru11}. 

Here, we focus on extending our previously reported work 
\cite{stru11}, using a more accurate sampling of the parameters 
space of our MHD model. Therefore, for the period of
2002-2007, which covers both V1 \& V2 crossings, we report a rich set of 
numerical simulations that include several strengths and orientations of the
ISMF, many SW conditions at different epochs that describe its anisotropy, and a fine 3D grid. 
Our primary goal is to fit V1, V2, and IBEX observations simulatneously,
which should provide an accurate measurement of the unperturbed ISMF
vector at the actual Galactic position of our solar system. A comparison of 
our findings with the most recent reports on the ISMF vector is also provided.
\section{Model and method}
\begin{figure}
\centering
 \includegraphics[width=7.cm]{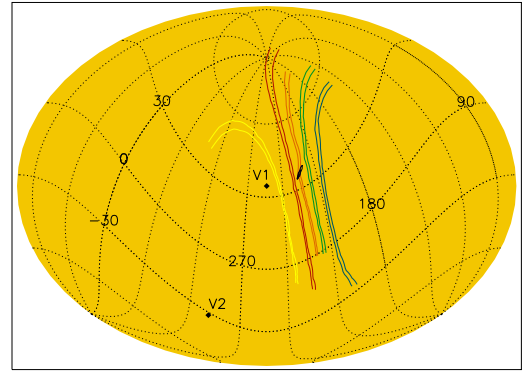}
\caption{Contours of regions in ecliptic coordinates of ISMF orientations where the MHD model reproduces the TS crossing for V2 on 2007 August 30 for ISMF strengths equal to (blue) 1.8, (green) 2.0, (orange) 2.2,
 (red) 2.4, and (yellow) 3 $\mu$G (contours are shown from right to left in increasing strength). For the IBEX ribbon, the small black
 contour shows  $|\mathbf{B}\cdot \mathbf{R}|<0.025$ obtained for $2.2 \mu$G as the best fit.  
We note that the $\mathbf{B}\cdot\mathbf{R}=0$ condition is not sensitive
 to the ISMF strength and only the best solution is shown.}
\end{figure}
We use the 3D MHD time-dependent model in which the effects of the
interplanetary magnetic field and the anisotropic solar wind are
simulated. {We solve the set of MHD equations with a
constant flux of hydrogen \citep{rb02}. This assumption is validated
afterward by comparing our final results to independent models that use a
full kinetic description for neutrals (e.g., section 3) \citep{heer11}. 
Furthermore, as a lever to uncover the ISMF vector, the TS asymmetry was 
shown to be unchanged by the treatment of neutrals, provided the density
level at that position is consistent with observations
\citep{alo11,izm03}. 
In our calculations the coordinate system is Sun-centered, and the x-y plane contains the ISMF
$\mathbf{B}_{\mathbf{\infty}}$ and velocity
$\mathbf{V}_{\mathbf{\infty}}$ vectors, with the latter parallel to the
x-axis direction.
At the outer boundary we use an LISM with 
plasma density $n_{\infty}=0.095$ cm$^{-3}$ and velocity 
\mbox{$V_{\mathrm{\infty}} =$ 26.4 km/s}, flowing from the He 
direction $(255.4^{\circ}, 5.2^{\circ})$ in ecliptic coordinates and 
temperature \mbox{$T_{\mathrm{\infty}} =$ 6400 K} {(hereafter called
Ulysses-based He parameters \citep{wit04})}. 
To have the neutral H number density at the TS consistent with measurements, 
we choose \mbox{$n^{\mathrm{\infty}}_\mathrm{H} =$ 0.11
cm$^{-3}$}, which is still in the range $0.1 - 0.2\, cm^{-3}$ indicated
by \cite{izm03} for the LISM}. The unperturbed ISMF strength varies
within the limit 1.5 to 3.0$\, \mu$G. 
The orientation of the ISMF in space is defined by two angles: an 
inclination angle $\alpha$ (between inflow $\mathbf{V}_{\mathbf{\infty}}$
and magnetic field $\mathbf{B}_{\mathbf{\infty}}$ vectors) 
which varies from $0^{\circ}$ to $90^{\circ}$, and a
deviation angle $\beta$ which varies from $-90^{\circ}$ to $90^{\circ}$ and 
measures the angle of the x-y plane from the hydrogen deflection plane (HDP) \citep{lal10}. 

The anisotropy of the 
solar wind flow is included using slow and fast 
wind sectors that are estimated from yearly maps of observed 
interplanetary scintillations. According to these observations, 
for the period of 2001-2005, the slow SW cone was $\sim 56\pm 6^\circ$, while for 
the period of 2006-2008, the cone
was almost $\sim36\pm 6^\circ$. During the intermediate period of 2005-2006, the 
slow SW sector was assumed to shrink linearly between the two values \citep{tok10}.

\begin{figure}
\centering
 \includegraphics[width=7.5cm]{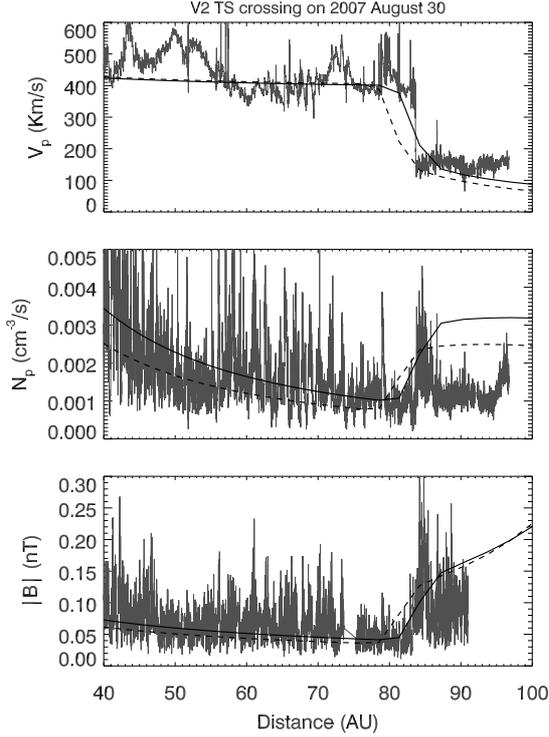}
\caption{ Comparison of the  V2 plasma and field measurements (gray) and MHD model (black solid) 
for simulation parameters $\alpha=42^\circ$, $\beta=0^\circ$, and ISMF strength $2.2\, \mu$G 
{that correspond to the best solution obtained in Figure 1 for the
 Ulysses He parameters. We also show MHD model (black dashed) 
for simulation parameters $\alpha=45^\circ$, $\beta=0^\circ$, and ISMF strength $3.0\, \mu$G 
 that correspond to a  solution recently proposed when using IBEX-based He parameters
 \citep{mcco12,bzo12}. The SW speed 
is $\sim 420$km/s}. From top to bottom, we show the 
plasma speed, density, and magnetic field strength.}
\end{figure}

In addition to the two values assumed for the slow SW extent, we also
considered several speeds at 1AU in the range of 420 to 500 km/s for the 
slow component in order to cover most values thus far measured by the 
plasma detector on V2 and by the {\it Advanced Composition Exoplorer} (ACE), 
an Earth-orbiting satellite. For the fast SW component, a factor 
$\sim 1.9$ times the slow speed value is assumed \citep{tok10}. For 
the TS crossing time reported so far for V1 \& V2,
the slow solar wind plasma density varies in the 
range of $n_\mathrm{SW}\sim 4.5-6.5 \, {\rm cm^{-3}}$ , but was fixed 
at $ n_\mathrm{SW}=5.2\, cm^{-3}$ so that the corresponding SW dynamic 
pressure (in the range $1.4-2.1\, $nPa at 1AU) follows the square of the SW speed variation.  
The solar equator is tilted $7.25^{\circ}$ with respect to the ecliptic plane. 
The ecliptic longitude of the Sun's ascending node is $75.77^{\circ}$
(J2000). The Parker's spiral model, which is assumed for the interplanetary 
magnetic field, has a radial component equal to 35.5$\, \mu$G at 1 AU.
 
In order to compare our MHD simulations to IBEX ribbon observations, we use the
locations of enhanced emission of ENA fluxes as reported by
\cite{funs09}. These locations are defined by ecliptic latitudes and 
longitudes at which we compute the $\mathbf{B}\cdot \mathbf{R}=0$ condition of perpendicularity
between the line of sight of the observation and the magnetic field vector
in the region outside the heliopause where the ISMF has its maximum
strength \citep{heer10,stru11}. It is important 
to stress that modeling of the ENAs flux requires kinetic models for 
neutrals and is outside the scope of our study \citep{heer10}. 

\begin{figure}
\centering
 \includegraphics[width=7.5cm]{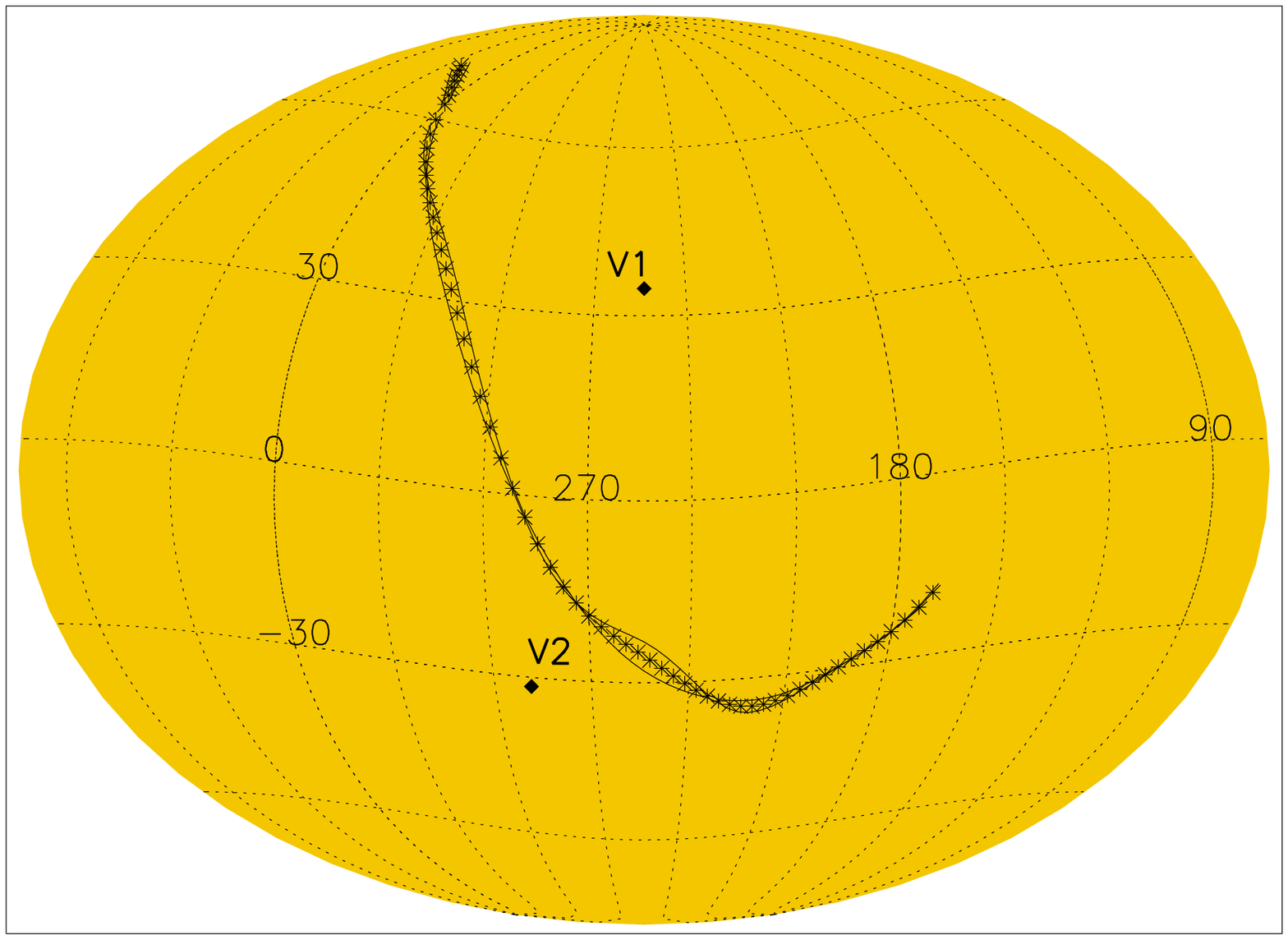}
 \includegraphics[width=7.5cm]{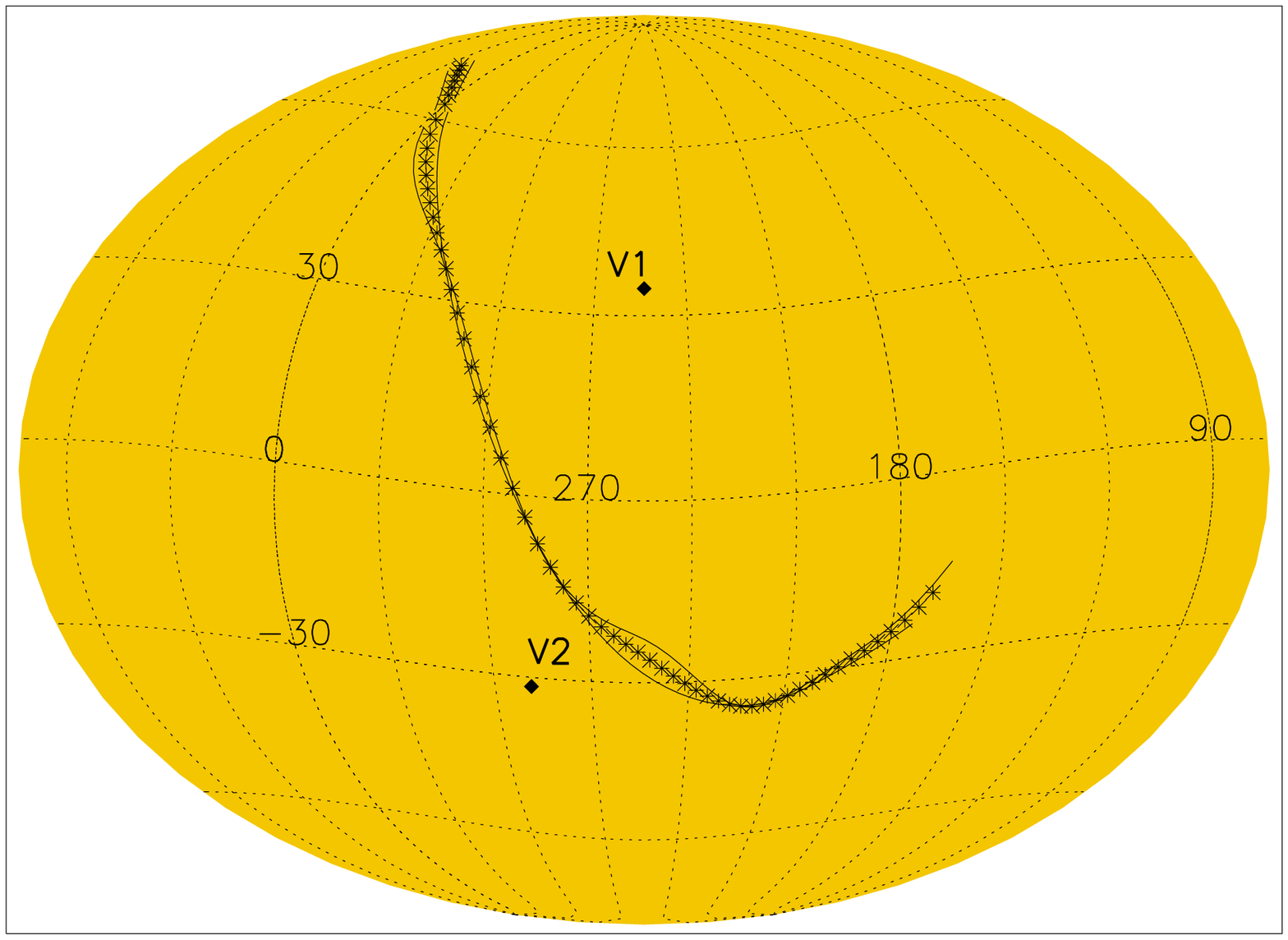}
\caption{{\bf (Top)}. Comparison of the angular positions (ecliptic
 coordinates) of computed (solid line) and observed (stars) IBEX ribbon
 estimated for the best fit derived in Figure 1 with SW parameters that
 best correspond to V2 TS-crossing time in 2007 (SW speed $\sim 420$km/s). {\bf (Bottom)}. Same but for SW
 parameters that best correspond to the V1 TS-crossing time in 2004 (SW speed $\sim 466$km/s). }
\end{figure}

In a first step, we simultaneously target the IBEX observation and the V2 TS 
crossing for which both plasma and field measurements exist. The angular
grid for ISMF assumes a five degree regular step and strengths varying in
the set $1.5$, $1.8$, $2.0$, $2.4$, and $3.0$ $\mu$G. After an approximate solution 
is derived for the strength and orientation of the ISMF vector, we
consider a finer grid ($\sim $2 degrees step) around that solution plus an
additional value for the strength. Next, we consider the V1 TS 
crossing in 2004, using those SW speed and 
density values that are consistent with V2 plasma measurements obtained at 
the time of the V1 crossing and with ACE measurements conducted at 1 AU a 
year or so before its propagation to the V1 position. The approach
described above should provide an ISMF vector consistent with all observations. 
\section{Results}
Following the method described above, our first step is to conduct a 
sensitivity study by comparing our MHD simulations with the V2 TS
crossing on 2007 August 30 and the IBEX ribbon observation. In 
Figure 1, we compare ecliptic latitudes/longitudes regions
for which the MHD model fits the plasma parameters distribution
during the V2 TS crossing for different strengths of the ISMF. We note 
that the ensemble of solutions is very sensitive to the field strength
and orientation. However, where the MHD model
generally recovers the IBEX ribbon, the space regions are not very sensitive to the field
strength, a finding that is consistent with previous studies
\citep{stru11,heer11}. It appears that while the IBEX ribbon strongly constrains
the ISMF orientation, it is less effective on the strength, which is better constrained by
the V2 TS crossing. This means that we need both data sets to uniquely
derive the ISMF strength and orientation. This statement is crucial for
understanding the difference between our results and studies using the
only IBEX ribbon data analysis. For clarity, a  detailled comparison
with the recent results of Mccomas et al. (2012) is provided at the end
of this section as a reference. 

Following the proposed step-by step analysis, our parametric study shows that our MHD
model reproduces the position of the IBEX ribbon and the TS crossing
conditions for V2 at $\sim 83.7$ AU, when the ISMF magnitude
is in the range $2.0-2.4\, \mu G$ and is pointing from a direction that 
corresponds to MHD space parameters $\alpha \sim 40-45^\circ$ and 
$\beta\sim(-5)-5^\circ$. This solution was obtained for a slow solar wind sector of
$36^{\circ}$ with a density of 
$\sim 5.2\, {\rm cm^{-3}}$ and a speed of $\sim 420$ km/s at 1 AU, consistent
with measurements obtained respectively by V2 and by ACE before SW
propagation to the TS crossing. To conclude this first step, we consider a finer grid in the ISMF space
parameters and an additional strength of $2.2\, \mu$G as revealed from the
previous analysis. As shown in Figure 1, the best fit is obtained when the
IBEX solution crosses the V2 TS solution for a ISMF vector with a strength of
$2.2\pm0.1\, \mu$G and pointing from ecliptic coordinates $(224, 36)\pm3^\circ$. This
solution corresponds to space parameters $\alpha\sim 42\pm2^\circ$ and 
$\beta\sim 0.0\pm 2^\circ$ of our MHD simulation. The fit to the 
plasma parameters during the V2 TS crossing is shown in Figure 2 (solid line), and the fit corresponding  
to the IBEX ribbon is shown at the top of Figure 3. 

The second step in our approach is to determine whether given the SW conditions 
that best correspond to the TS crossing on 2004 December 16, and with the 
ISMF vector derived in step 1, our MHD model provides a good fit to the position of the 
TS as measured by V1 in the interplanetary magnetic field. As shown in
Figure 4 (solid line) a rather good fit is obtained with the same ISMF strength 
and orientation obtained from V2 and IBEX data analysis but for SW speed 
$\sim 466$ km/s and density $5.2\, {\rm cm^{-3}}$ (SW ram pressure $\sim
1.886$ nPa).  
In addition, for the V1 plasma conditions derived above, a good fit is also obtained for the IBEX
ribbon (Figure 3, bottom), confirming the weak sensitivity of the outer
heliosheath to SW conditions \citep{izm05}. However, our sensitivity analysis shows that for 
the V1 events thus far reported in mid-2002 and early 2003 with the corresponding 
noisy IMF data \citep{bur03}, the TS position could be 
in the range 85-94 AU. Apparently, predicted positions of the TS are at or a few AU 
ahead of the V1 position, yet still close to it. Because of large uncertainties in plasma 
and field parameters, our results cannot distinguish between the 
two scenarios proposed so far for the 2002/2003 events \citep{krim03,jok04,krim05}. 
Ultimately, detailed modeling using the ISMF vector proposed here will 
be much needed in the future in order to check the flux level and energy range of 
particles measured by V1.
\begin{figure}
\centering
 \includegraphics[width=7.5cm]{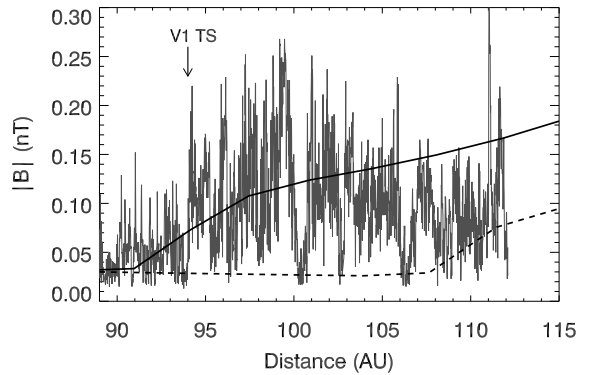}
\caption{ {Comparison of V1 field measurements (gray) and MHD
 model (solid line) for simulation parameters $\alpha=42^\circ$, $\beta=0^\circ$, and ISMF strength $2.2\, \mu$G 
that correspond to the best solution obtained in Figure 1 but with SW speed $\sim 466$km/s and Ulysses-based He parameters. We also show MHD model (black dashed) 
for simulation parameters $\alpha=45^\circ$, $\beta=0^\circ$, and ISMF strength $3.0\, \mu$G 
that correspond to the IBEX-based He parameters \citep{mcco12}.}}
\end{figure}

{To clarify why it is critical to use most of the available {in situ} measurements 
and the remote observations to properly constrain the ISMF vector, we decided to 
compare our findings to the recent report on the ISMF strength and orientation 
derived from the only IBEX data set, including the
new ISM He orientation, speed, and temperature (hereafter called
IBEX-based He parameters) \citep{bzo12,mcco12}. Our simulations show that 
the corresponding MHD model (dashed line) does 
not correctly fit the {\it in situ} measurements obtained by V2 in
2007 (Figure 2) and V1 in 2004 (Figure 4). Because the obtained (dashed) curves
fall on opposite sides of the TS position respectively observed by V1
and V2, we do not see how a stronger ISMF ($\geq 3.0\, \mu$G) would help better fit both 
Voyager measurements. Because the Voyagers' TS crossing and IBEX He data were obtained at different
epochs, the obtained misfit may have several readings. First, our
results cast doubt on the recently proposed strong ISMF
solutions, and the consequent conclusions regarding the
non-existence of the ISM bow shock \citep{mcco12}. Second, for the
newly assumed He parameters, other ISMF vector solutions may exist and should be tested in the
future following the sensitivity approach sketched in Figure 1. 
Finally, if a stronger ISMF is assumed, the misfit obtained here with the Voyager
data could also indicate that either the IBEX He data should be critically 
re-analyzed \citep{mob12}, or the IBEX He parameters reflect a real
change in the ISM flow that occurred after the V2 TS crossing in 2007.}

To further verify and extend the scope of our results, we tried to check
 their consistency. First, we point out that our analysis put the ISMF orientation very close to the observed HDP plane as reported by Lallement et al. (2010). Furthermore, both the orientation and
 strength derived here are consistent with the values obtained from an independent analysis
 constrained by the same HDP plane and a full kinetic treatment of
 neutrals \citep{heer11}. This consistency between independent studies 
 justifies our assumption of a constant H flux. In addition, we tried to compare 
the ISMF vector derived here to existing estimates obtained by other different 
techniques. Table 1 summarizes most of the reports on the ISMF
derived from the V1 \& V2 TS crossing and IBEX ribbon interpretation. 
Independently, Lyman-$\alpha$ sky-background observation, rotation measure 
and dispersion measure of radio pulsars, along with starlight polarization, provided 
an estimation of the ISMF vector in the solar system neighborhood \citep {rbg08,sal10,fri10}. As 
shown in Figure 5, the comparison of Galactic coordinates for the different estimates 
shows good convergence between the different techniques, lending credence to 
our results and the method utilized to obtain them. 
\begin{figure}
\centering
 \includegraphics[width=7.5cm]{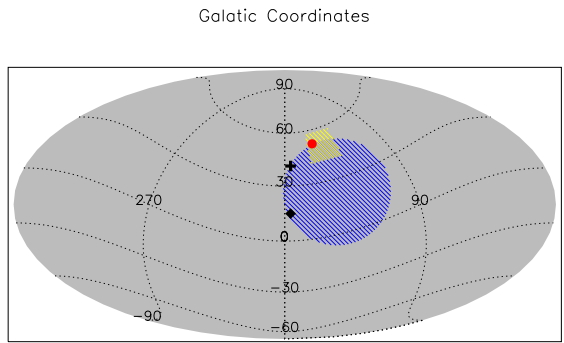}\caption{Galactic coordinates of ISMF. (Yellow, hashed square) Sky background \lya\ solutions (37$\pm14^\circ$, 49.5$\pm8.5^\circ$). (Red, small circle) Solutions from {\it Voyager} \& {\it IBEX} data analysis (28, 52)$\pm 3^\circ$ (this work).  (Blue, large hashed circle) Starlight polarization solutions (38, 23)$\pm35^\circ$ \citep{fri10}. Big cross shows the solution of the Faraday rotation measure and dispersion measure (5$^\circ$, 41$^\circ$) for 7 pulsars \citep{sal10}. Filled diamond is upwind direction. Angles increase counterclockwise. }
\end{figure}
\section{ Summary and conclusions} 

We have shown that the asymmetries observed in the sky background \lya\ emission 
deep in the heliosphere, in the TS position as measured by V1
and V2, and in the IBEX ribbon of ENAs, can all be explained with a rather
weak $ |\mathbf{B}_{\mathbf{\infty}}|\sim 2.2\pm 0.1\, \mu$G ISMF pointing from 
Galactic coordinates $(28, 52)\pm 3^\circ $. 
This direction makes an angle of $\sim 42^\circ$ with the Ulysses helium flow in
agreement with the value first found by \cite{bpr00}. The ISMF vector
derived here lies in the HDP plane \citep{lal10}, consistent with an
independent analysis of the SOHO/SWAN hydrogen flow deflection angle
\citep{heer11}. We have also shown that the 
magnetohydrodynamic modeling of the only V1 and V2 {\it in situ}
measurements or IBEX data set does 
not lead to unique values of the ISMF strength and orientation as 
previously claimed (e.g., Figure 1). The asymmetry thus far reported for
the TS shape is not only the result of the ISMF effect but also of the SW 
anisotropy and time variation.  

{Our solution for the ISMF vector indicates a weak field that is
consistent with all reported {\it in situ}
measurements and remote observations. For the Ulysses-based He
parameters thus far assumed for the interstellar flow, a bow shock must
exist. By contrast, using a strong magnetic field
($\geq 3.0\, \mu$G) making $\sim 45^{\circ}$ from the new IBEX-based He
flow leads to a plasma configuration that is not consistent with the V2 and V1
TS crossing. Our results cast doubt on the strong ISMF solutions
recently proposed by McComas et al. (2012). Yet for the newly 
proposed interstellar He parameters, further simulations are required 
before one might conclude whether the interstellar bow shock does disappear 
under those assumptions.}

The ISMF derived here, at the 
actual Galactic position of the Sun, is in the range $\sim 70-80^{\circ}$ from 
the different estimations of the large-scale (kpc) Galactic magnetic
field (e.g., Figure 5). This finding is consistent 
with previous studies of rotation measure and dispersion measure of
radio pulsars and from starlight polarization \citep{sal10,fri10}. 

\begin{acknowledgements}
LBJ acknowledges support from Universit\'e Pierre et Marie Curie (UPMC) and the Centre National de la Recherche Scientifique (CNRS) in France. RR acknowledges support from Foundation for Polish Science, contract Focus F3/07/2011. Authors acknowledge support from LEA ASTRO-PF. 
\end{acknowledgements}

\end{document}